\documentclass[aps,prd,showpacs,nofootinbib]{revtex4}

\usepackage{graphicx}
\usepackage{color}

\def\beq{\begin{eqnarray}}
\def\eeq{\end{eqnarray}}

\begin{document}

\title{Going beyond perturbation theory: Parametric Perturbation Theory}
\author{Paolo Amore}\email{paolo@ucol.mx} 
\affiliation{Facultad de Ciencias, Universidad de Colima, \\
Bernal D\'{\i}az del Castillo 340, Colima, Colima, Mexico} 

\begin{abstract}    
We devise a {\sl non--perturbative} method, called {\sl Parametric Perturbation Theory} (PPT), which is
alternative to the ordinary perturbation theory. The method relies on a principle of simplicity for the 
observable solutions, which are constrained to be linear in a certain (unphysical) parameter. The perturbative
expansion is carried out in this parameter and not in the physical coupling (as in ordinary perturbation theory).
We provide a number of nontrivial examples, where our method is capable to resum the divergent perturbative series,
{\sl extract} the leading asymptotic (strong coupling) behavior and {\sl predict} with high accuracy 
the coefficients of the perturbative series. In the case of a zero dimensional field theory we prove that PPT
can be used to provide the imaginary part of the solution, when the problem is analytically continued to 
negative couplings. In the case of a $\phi^4$ lattice model $1+1$ and of elastic theory we have shown that
the observables resummed with PPT display a branch point at a finite value of the coupling, signaling the transition
from a stable to a metastable state. We have also applied the method to the prediction of the virial
coefficients for a hard sphere gas in two and three dimensions; in this example we have also found that the solution 
resummed with PPT has a singularity at finite density. Predictions for the unknown virial coefficients are made.
\end{abstract}
\maketitle


\section{Introduction}
\label{intro}

For the majority of the problems in Physics no exact analytical solution is known
and it is a common procedure to resort to Perturbation Theory (PT)~\cite{PT1,PT2,PT3,PT4}.
The fundamental idea behind perturbation theory is that, if a given problem is solvable
for a particular value of a parameter, then one can obtain analytical approximations to 
the solution in a close neighborhood of this value, by Taylor expanding in that parameter. 
Calling $g$ this parameter and $g_0$ the value for which an exact analytical solution 
is known, then the results obtained with perturbation theory will be, to a given order, 
polynomials in $(g-g_0)$. The perturbative series obtained by considering all the terms 
of the  expansion will in general have a finite radius of convergence, $r$, 
which in some cases could even be zero and therefore the series would be divergent for 
all $g \neq 0$. Actually divergent series are usually expected from the application
of PT to quantum field theory, as first observed by Dyson in the case of Quantum 
Electrodynamics (QED)\cite{Dyson52}. Another well--known example of divergent series 
is given by the quantum anharmonic oscillator, whose perturbative coefficients for the 
energy of the ground state have been calculated by Bender and Wu in \cite{Bender} and 
proved to have a factorial growth.

Although the pertubative series provide in many cases the only systematic approach
to the solution of a problem, they are not always useful, since they are confined to 
a restricted region for the physical parameters, $|g-g_0| < r$. Outside this region,
the physics becomes {\sl nonperturbative} and cannot be described directly in terms 
of the original series. 
For quite a long time physicists have been interested into finding a bridge from the 
perturbative  to the non--perturbative region. Several methods have been developed which
allow to extract the non--perturbative behavior from the perturbative series: among such 
methods we would like to mention the Borel and Pad\'e approximants\cite{BenderOrzag, PT1} 
and nonlinear  transformations \cite{Weniger89}.

Methods which are alternative to PT should retain on one hand the ability to provide a sistematic
analytical approximation to a given problem, and on the other hand they should remain valid even 
in the nonperturbative region, never leading to divergent series. Over the years new ideas 
have allowed to devise methods which comply with these requirements.  The Linear Delta 
Expansion (LDE) \cite{lde} and the Variational Perturbation Theory (VPT) \cite{Klei04} are two
examples of  non--perturbative methods. 
Roughly speaking these methods work by introducing in a problem an artificial parameter and 
turn the original problem into a new one with a modified perturbation. The optimization 
of the ``perturbative'' results to a given order with respect to the artificial parameter is 
usually obtained through the {\sl Principle of Minimal Sensitivity} (PMS)\cite{Ste81} and leads 
to expressions which are non--polynomials in the physical parameters and therefore non--perturbative. 

In this paper we will explore a new a path, which is also described in shorter letter: 
the method that we have devised, which we have called 
Parametric Perturbation Theory (PPT) method, is based on few simple ideas. The first one, which 
we will refer to as {\sl Principle of Absolute Simplicity} (PAS), is that we do not want to calculate 
the observable (energy, frequency, etc.) directly as a polynomial  in the physical coupling $g$, 
as done in PT, but that this observable should have the simplest possible form (linear) in a given 
unphysical parameter $\varrho$; the second idea is that the perturbation theory must be carried
out in $\varrho$ and that the functional relation $g = g(\varrho)$ must comply with the Principle 
of Absolute Simplicity to the order to which the calculation is done. This will allow to determine
the relation between $g$ and $\rho$ and in turn to obtain the observable as a parametric function
of $\varrho$. 

The paper is organized as follows: in Section \ref{sec1} we develop the Parametric Perturbation Theory 
for a problem of nonlinear oscillations in classical mechanics and show that at finite order it provides  
extremely accurate approximations for the frequencies and for the solutions of the problem; in Section 
\ref{sec2} we show that the PPT approach can be applied directly on the perturbative series and discuss 
the performance of our method in the case of non trivial examples, with divergent perturbative series.
Finally, in Section \ref{concl} we draw our conclusions.

\section{The method}
\label{sec1}

Consider a model which depends on a parameter $g$, and which is solvable when $g=0$. The application of 
PT to this problem to a finite order yields a polynomial in $g$. Calling $r$ the radius of convergence of
the perturbative series, the direct use of PT must be restricted to $|g|<r$, as previously discussed.
However, the misbehavior of the perturbative series for a physical observable $\mathcal{O}$ is the result of 
having expanded in a parameter, $g$, which is not optimal. If one knew the exact solution to the problem, 
i.e. $\mathcal{O} = f(g)$, then  this solution could be considered as a polynomial of order one in the variable 
$\varrho = f(g)$. Although this observation by itself cannot be used as a constructive principle, we may adopt
the philosophy that the {\sl perturbative series for the observable can be simpler and  convergent in all the 
domain, if it is cast in terms of a suitable parameter $\varrho$}. 

Only if such parameter, by luck or ability, turns out to be the $\varrho = f(g)$ discussed above, 
the exact solution is obtained. The goal, therefore, is to progressively build this parameter $\varrho$ 
to yield an expression for $\mathcal{O}$ as simple as possible. 
In this framework the perturbative expansion is carried out in $\varrho$ and all the physical quantities 
in the problem are expressed as functions of $\varrho$. In particular we have now that $g = g(\varrho)$. 
While the ordinary perturbation theory works by calculating the contributions to higher orders in $g$, 
each term of higher order refining the result to lower order, the approach approach is the opposite: we carry 
out a perturbative  calculation in $\varrho$, and then determine order by order the form of $g = g(\varrho)$ so 
that the observable $\mathcal{O}(\varrho)$ can be a order one polynomial in $\varrho$. This is in essence 
the {\sl Principle of Absolute Simplicity}.

Having given the general ideas of the method we proceed to examine its implementation
in a concrete problem. We consider the classical nonlinear oscillations of a point mass described by the 
equation (Duffing equation)
\begin{equation}
\frac{d^2x}{dt^2} + x(t) =  - g x^3(t) \ .
\label{eq:1}
\end{equation}

The Lindstedt-Poincar\'e method can be used to obtain a perturbative expansion of the 
squared frequency $\Omega^2$ of the oscillations in powers of $g$~\cite{PT1,PT3,PT4}. The method works by defining 
an absolute time scale, independent of $g$ and by then fixing the coefficients of the expansion of $\Omega^2$ 
so that the secular terms in the expansion are eliminated at each order. Working through order $(gA^2)^5$ one finds
\begin{equation}
\Omega^2 \approx 1+\frac{3 g A^2}{4}-\frac{3 g^2 A^4}{128}+\frac{9 g^3 A^6}{512}
-\frac{1779 g^4 A^8}{131072} + O\left[(g A^2)^5\right] \ ,
\label{eq:2}
 \end{equation}
$A$ being the amplitude of oscillations\footnote{The radius of convergence of the perturbative series 
in this case is $\bar{g} = 1/A^2$.}. 

We now proceed to implement our method. The first step is to define a functional relation 
between $g$ and the perturbative parameter of the expansion, $\varrho$. 
For example, we choose
\begin{equation}
g(\varrho) = \varrho \ \frac{1+\sum_{n=1}^{\bar{N}} c_n \varrho^{n}}{1+\sum_{n=1}^{\bar{N}} d_n \varrho^{n}} \ ,
\label{eq:3}
\end{equation}
where the coefficients $c_n$ and $d_n$ are unknown constants to be later determined.
The parameter $\bar{N}$ is related to the order to which the calculation is performed, since the 
number of coefficients $c_i$ and $d_i$ needs to match the number of conditions available at a given order. 

The reader may wonder the reason of the particular choice made in eq.~(\ref{eq:3}): the functional relation
between $g$ and $\varrho$ can be more general than eq.~(\ref{eq:3}), although it is not completely arbitrary
since it must reproduce all the terms in the perturbative expansion when the relation between $g$ and $\varrho$
is inverted. The choice made here takes into account this fact and also the leading asymptotic behavior 
of the frequency, $\lim_{g\rightarrow \infty} \Omega^2 \propto g$.\footnote{We are assuming that the denonimator
of eq.(\ref{eq:3}) does not have zeroes for $\varrho >0$ and that therefore $g = \infty$ is reached for
$\varrho = \infty$.}

We now follow the standard procedure of the LP method and introduce an absolute time
$\tau \equiv \Omega(\varrho) \ t$. Applying the PAS we choose this relation to be
\begin{equation}
\Omega^2(\rho) = \alpha_1 + \alpha_2 \varrho \ ,
\label{eq:4}
\end{equation}
where $\alpha_{1,2}$ are coefficients to be determined. We also expand the solution as
\begin{equation}
y(\tau) = \sum_{n=0}^\infty y_n(\tau) \ \left(\frac{g}{\varrho}-1\right)^n \ .
\end{equation}

The reader familiar with the LP method will recognize the profoundly different character of the present 
approach: in the LP method the observable, i.e. $\Omega^2$, is expressed as a series in powers of $g$, 
whose coefficients are determined by the condition that secular terms are eliminated at each order; here 
we impose that $\Omega^2$ has the simplest possible form when expressed in terms of $\varrho$, and we let 
$g(\varrho)$ to contain arbitrary powers of $\varrho$. In detail, we transform the original equation into
the new equation
\begin{equation}
\Omega^2(\varrho) \ \frac{d^2y}{d\tau^2} + y(\tau) = - g(\varrho) \ y^3(\tau) \ .
\label{eq:5}
\end{equation}

For sake of simplicity we will limit ourselves to work through order $\varrho^3$ and solve the differential 
equations resulting at each order in $\varrho$. The elimination of the secular term to order one yields  
$\alpha_1 = 3 A^2/4$, as in standard LP method  ($\alpha_0 = 1$). 
The solutions to order $0$ and $1$ are $y_0(\tau) = A \cos\tau$ and 
$y_1(\tau) = \frac{A^3}{32 (c_1-d_1)} \ (-\cos\tau +\cos 3\tau)$.

The elimination of the secular term to second order provides the condition
\begin{equation}
 d_1 = c_1 - \frac{A^2}{32}
\end{equation}
and the solution
\begin{equation}
y_2(\tau) = \left( 23 A - \frac{32 c_1}{A} \right) \cos\tau  - 
\left( 24 A - \frac{32 c_1}{A} \right) \cos 3\tau  \ .
\end{equation}

\begin{figure}
\begin{center}
\bigskip\bigskip\bigskip
\includegraphics[width=14cm]{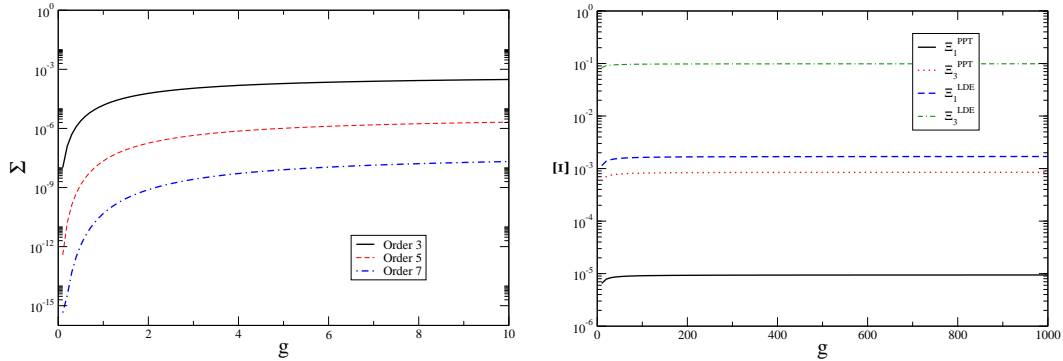}
\caption{(color online) 
Left panel: Error over the square frequency, $\Sigma = 1-\Omega^2_{approx}/\Omega^2_{exact}$, calculated to 
order 3,5 and 7 (solid, dashed and dot-dashed curves); Right panel: Error over the first two Fourier coefficients calculated to order 3 using PPT 
and the LDE approach of \cite{Amore07}.}
\label{Fig_1}
\end{center}
\end{figure}

Finally, one can determine $c_1$ to cancel the secular term to third order, 
$c_1 = \frac{23}{32} A^2$, and correspondingly $d_1  = \frac{11}{16} A^2$.
The solution to third order thus reads
\begin{equation}
y_3(\tau) =5 A \cos \tau - 4 A \cos 3\tau -2 A \cos 5 \tau + A \cos 7 \tau \ .
\end{equation}

To order $\varrho^3$ we therefore find
\begin{eqnarray}
g &\approx& \varrho \ \frac{32+23 \varrho A^2  }{32+22 \varrho A^2}  \ ,
\end{eqnarray}
which can be inverted and used to express the frequency directly in terms of $g$:
\begin{eqnarray}
\Omega^2 &\approx& 1 + \frac{3}{4} \varrho A^2 = \frac{11}{23} + \frac{33}{92} g A^2 
+ \frac{3}{92} \sqrt{\left(121 g A^2+384\right) g A^2+256} \ .
\end{eqnarray}

Notice that this expression is non--perturbative in $g A^2$ and that it provides
a maximum error $\Sigma = \lim_{g A^2 \rightarrow \infty} 
\left(1-\Omega^2/\Omega^2_{exact} \right)  \approx 5.27 \times 10^{-4}$.
This error is much smaller that then the one obtained to third order in \cite{Amore03} using the Linear Delta 
Expansion, i.e. $\Omega^2_{LDE} = \frac{69 g A^4+192 g A^2+128}{96 g A^2+128}$ ( in which case one has
$\Sigma_{LDE} = \lim_{g A^2 \rightarrow \infty} \left(1-\Omega^2_{LDE}/\Omega^2_{exact} \right) 
\approx 1.36 \times 10^{-2}$).
 
At the same time PPT provides highly accurate estimates for the Fourier coefficients of the solution, $c_n$. 
In the right panel of fig.\ref{Fig_1} we have plotted the error defined 
as $\Xi_n \equiv \left| c_n^{approx}/c_n^{exact}-1\right|$ as function of $g$ and compared 
our results with the already excellent results of \cite{Amore07}. 

At this point the reader should recognize that, if one is interested only in frequency and not in the solution, 
it is possible to determine the coefficients $c_n$ and $d_n$ to a given order with a  minimal effort directly 
from the coefficients of the perturbative series.
The procedure consists of first substituting $g=g(\varrho)$ inside the perturbative series, 
and then expanding around $\varrho=0$: the unknown coefficients $c_n$ and $d_n$ are then 
used to suppress the nonlinear behavior in $\varrho$ inside $\Omega^2$. 
Following this simple procedure we have produced the result to order $5$ and $7$  in the left panel
of Fig.\ref{Fig_1}.

\section{Resummation of perturbative series}
\label{sec2}

In many cases the perturbative results for a given problem are known only to a finite order since the calculation
of each higher order involves increasing technical difficulties. For example, the calculation of observables 
in Quantum Field Theory at a given order in PT requires to take into account a number of  Feynman diagrams
which rapidly grows with the order of the perturbation, the calculation of the higher order (multiloop) diagrams
being more and more challenging. 

For this reason it is desirable to have a procedure which, using only a finite number of perturbative coefficients,
may extract the essential physical behavior of the solution and possibly predict a number of unknown 
perturbative coefficients. We will show now that this result can be efficiently achieved using our method. 

\subsection{Anharmonic oscillator}
\label{sub1}

Consider the quantum anharmonic oscillator
\beq
\hat{H} = \frac{\hat{p}^2}{2} + \frac{x^2}{2} + g x^4 \ .
\label{sub1_1}
\eeq
The series obtained with perturbation theory for this problem is divergent and its coefficients behave as 
\beq
b_n \approx (-1)^{n+1} \sqrt{6/\pi^3} \ \Gamma(n+1/2) 3^n
\label{sub1_2}
\eeq 
for $n\rightarrow \infty$ as shown in ~\cite{Bender}. These coefficients can be also obtained exactly, using the
recursion relations given by Bender and Wu in \cite{Bender}.
The resummation of this perturbative series has been considered by several authors, using different 
techniques\cite{PT1,PT2,LMSW69,WCV93,Ivanov95,Weni96,CWBS96,SG98,Home,Nuñez,RB06,KJ95}. 

We will follow the philosophy of our method and define a functional relation between the physical coupling $g$ and
the unphysical parameter $\varrho$. In principle this relation can be expressed by mean of an arbitrary function, but 
it can also use information coming from the strong coupling regime, where the energy goes like $E_0 \propto g^{1/3}$ 
as $g\rightarrow \infty$. For example we can choose
\beq
g(\varrho) = \varrho \ \left[ \frac{1+\sum_{n=1}^{\bar{N}+1} c_n \varrho^{n}}{1+\sum_{n=1}^{\bar{N}} d_n \varrho^{n}}  
\right]^2 ,
\label{sub1_3}
\eeq
which has the correct asymptotic behavior  for $g\rightarrow \infty$, provided that the denonimator does not vanish
for $\varrho >0$. The unknown coefficients $c_n$ and $d_n$ in this expression will be determined so that the 
ground state energy is linear in the unphysical parameter, as required by the PAS, i.e. 
\beq
E_0 = b_0 + b_1 \varrho \ .
\label{sub1_4}
\eeq

Choosing $\bar{N} = 3$, corresponding to use only the first $6$ perturbative coefficients, 
we can fully determine the coefficients $c_n$ and $d_n$:
\beq
c_1 &=& \frac{3111725471}{109345364} \ , \ c_2 = \frac{292194444505}{1749525824} , 
c_3 = \frac{1136953355311}{6998103296} \nonumber \\
d_1 &=& \frac{730092771}{27336341} \ , \ d_2  = \frac{215945995035}{1749525824} \nonumber \ .
\eeq

Working to this order it is possible to extract the leading coefficient of the strong coupling series
\beq
\alpha_0 &=& \lim_{\varrho \rightarrow \infty} \frac{b_0+ b_1 \varrho}{g(\varrho)} = 
3 \left(\frac{215945995035}{2273906710622}\right)^{2/3} \approx 0.624458 \ ,
\label{sub1_5}
\eeq
which should be compared with the fairly precise results of \cite{PT1,Fernandez97,KJ95}, 
$\alpha_0 = 0.66798625915577710827096$.
In the opposite limit, $g\rightarrow 0$, the energy calculated with the PPT can be cast in terms of $g$ 
after inverting (\ref{sub1_3}):
\begin{widetext}
\beq
E_0 &\approx& \frac{1}{2} +\frac{3g}{4} -\frac{21 g^2}{8} + \frac{333 g^3}{16} 
-\frac{30885 g^4}{128} +\frac{916731 g^5}{256}-\frac{65518401 g^6}{1024} 
+\frac{9397438011180958461 g^7}{7166057775104}  \nonumber \\
&-&\frac{5778726063447202343420510691 g^8}{195893798965944254464}  
+\frac{120539916022637946813802592301161300277 g^9}{171360630026191984864597639168} \nonumber \\
&-&\frac{20343642843373228886854633070691634893375294933 \
g^{10}}{1171093154092705757431369973022851072} + O\left[g^{11}\right]
\eeq
\end{widetext}
which is correct up to order $g^6$. In Table \ref{table1} we compare the exact perturbative coefficients
going from the order $g^7$ to order $g^{10}$ with the approximate ones predicted by the PPT using $\bar{N}=3$.
The last column displays the error $\Sigma_n \equiv 100 \times \left| b_n^{[3,2]}/b_n^{exact}-1 \right|$.

\begin{widetext}
\begin{table}
\caption[t1]{Comparison of the exact perturbative coefficients for the anharmonic oscillator with the 
approximate coefficients obtained with PPT using $\bar{N} = 3$.}
\begin{tabular}{|c|c|c|c|}
\colrule
n & $b_n^{(exact)}$ & $b_n^{[3,2]}$  & error ($\%$) \\
\colrule
7 & $\frac{2723294673}{2048}$ & $\frac{9397438011180958461}{7166057775104}$ & $1.399$ \\
8 & $-\frac{1030495099053}{32768}$ & $-\frac{5778726063447202343420510691}{195893798965944254464}$ & $6.19$ \\
9 & $\frac{54626982511455}{65536}$ & $\frac{120539916022637946813802592301161300277}{171360630026191984864597639168}$ & 
$15.61$ \\
10 & $-\frac{6417007431590595}{262144}$ & 
$-\frac{20343642843373228886854633070691634893375294933}{1171093154092705757431369973022851072}$ & $29.03$ \\
\colrule
\end{tabular}
\label{table1}
\end{table}
\end{widetext}

The reader could argue that the quality of the results that we have obtained is due to having taken into 
account the exact asymptotic behavior of $E_0$ for $g\rightarrow \infty$. We will now show that our method 
allows one to optimize the asymptotic behavior of the approximate solution, even in the case where the exact 
behavior is unknown. 
We consider the approximations corresponding to 
\beq
g(\varrho) = \varrho \ \left[ \frac{1+\sum_{n=1}^{\bar{N}_u} c_n \varrho^{n}}{1+\sum_{n=1}^{\bar{N}_d} d_n \varrho^{n}}  \right]^2 ,
\eeq
fixing $\bar{N}_u+\bar{N}_d=5$ as in the previous case. We can compare the first coefficient predicted by our method, $b_7$,
using the different sets. These coefficients are displayed in the first row of Table \ref{table2}; the second row  displays the
error (in $\%$) with respect to the exact coefficient, shown in Table \ref{table1}. 
The set $[3,2]$  previously considered provides the lowest error and therefore selects the correct asymptotic behavior.

This example shows that, even in the unfortunate case where the asymptotic behavior of the energy is unknown, 
it is possible to extract some information on the strong coupling regime directly from the perturbative series, using
a limited number of perturbative coefficients.

\begin{widetext}
\begin{table}
\caption[t1]{Comparison of the exact perturbative coefficients for the anharmonic oscillator with the approximate coefficients
obtained with PPT using $\bar{N} = 3$.}
\begin{tabular}{|c|c|c|c|c|c|c|}
\colrule
  & $b_7^{[5,0]}$  & $b_7^{[4,1]}$ & $b_7^{[3,2]}$ & $b_7^{[2,3]}$ &  $b_7^{[1,4]}$ & $b_7^{[0,5]}$ \\
\colrule
$b_7$ & $\frac{141732231981}{131072}$ & $\frac{49825588453972797}{38645137408}$ & 
$\frac{9397438011180958461}{7166057775104}$ & $\frac{31966088112282317691}{24592236412928}$ &
$\frac{67034980866178137}{52602994688}$ & $\frac{134498076375}{131072}$ \\
error ($\%$) & $22.97$ & $3.13$ & $1.399$ & $2.299$ & $4.34$ & $29.58$ \\
\colrule
\end{tabular}
\label{table2}
\end{table}
\end{widetext}

\begin{figure}
\begin{center}
\bigskip\bigskip\bigskip
\includegraphics[width=7cm]{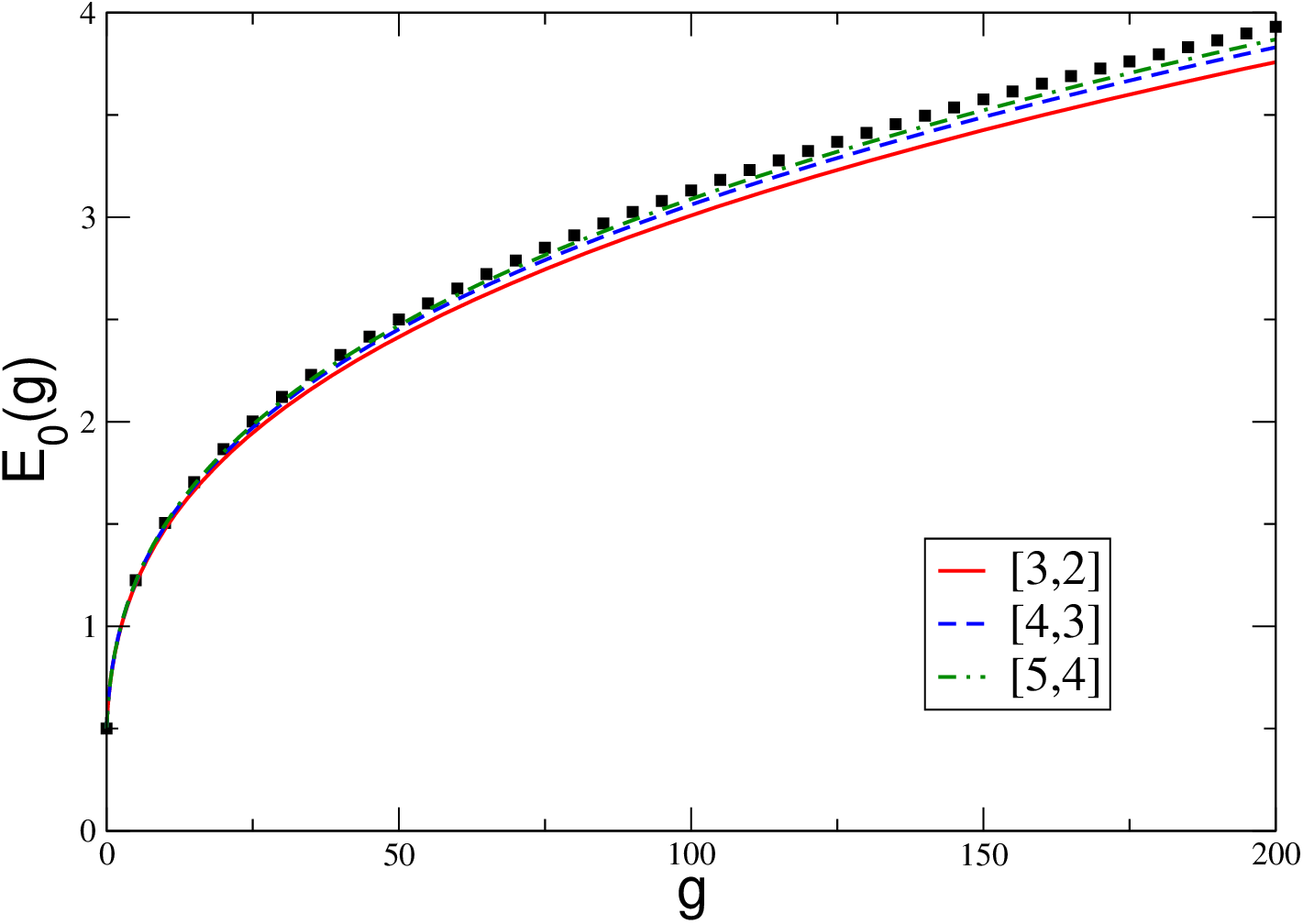}
\caption{(color online) Ground state energy of the anharmonic oscillator as a function
of $g$. The squares are numerical results, the curves correspond to the results obtained
with PPT to different orders.}
\label{Fig_2}
\end{center}
\end{figure}

In Fig.\ref{Fig_2} we have plotted the exact energy (numerical) as a function of $g$ (squares) and we
have compared it with the results of PPT applied to $3$ different orders, all reproducing the exact asymptotic
behavior. Our results approach the numerical result as $\bar{N}$ is increased.

\subsection{A {\cal PT} symmetric hamiltonian}

The complex hamiltonian 
\beq
\hat{H} = p^2+\frac{1}{4} x^2 + i g x^3 \ , 
\eeq
has been the first example of a {\cal PT} symmetric hamiltonian which has a 
completely real spectrum to be discovered. Bender and Dunne have studied 
in \cite{Bender98} the large order perturbative expansion of the ground state energy
of this hamiltonian finding that the coefficients of this series grow as
\beq
b_n \approx (-1)^{n+1} \frac{60^{n+1/2}}{(2 \pi)^{3/2}} \ \Gamma(n+1/2) \left[ 1 - O(1/n)\right] . 
\eeq
Table I of \cite{Bender98} contains the first 20 coefficients of the perturbative series. In a recent paper
Bender and Weniger \cite{BW01} have provided numerical evidence that the perturbative series for this
{\cal PT} symmetric hamiltonian is Stieltjes, using the first $193$ nonzero coefficients.

We will here use this model to obtain a further test of PPT. We assume the functional 
relation
\beq
g(\varrho) = \sqrt{\varrho \ \left[ \frac{1+\sum_{n=1}^{\bar{N}} c_n \varrho^{n}}{1+\sum_{n=1}^{\bar{M}} 
d_n \varrho^{n}} \right]} \ , 
\eeq
where the difference $\bar{N}-\bar{M}$ constrains the asymptotic behavior, which is not known
exactly in this case. Notice that the square root in the definition of $g$ is a consequence of the fact that the 
perturbative series contains only even powers of $g$ \cite{Bender98}. 
In Fig.\ref{Fig_3} we have compared the exact numerical results of the last column of
Table III of \cite{Bender98} with the calculation obtained with PPT using three different 
sets of $(\bar{N},\bar{M})$, which correspond to the same number of conditions~\footnote{Of course, when 
numerical results are not available, one can resort to the same approach followed for the anharmonic oscillator, 
selecting the optimal asymptotic behavior among those available.}.
Our results suggest that the asymptotic behavior of the energy is approximately 
$E_0 \propto \sqrt{g}$ for $g\rightarrow \infty$.

\begin{figure}
\begin{center}
\bigskip\bigskip\bigskip
\includegraphics[width=7cm]{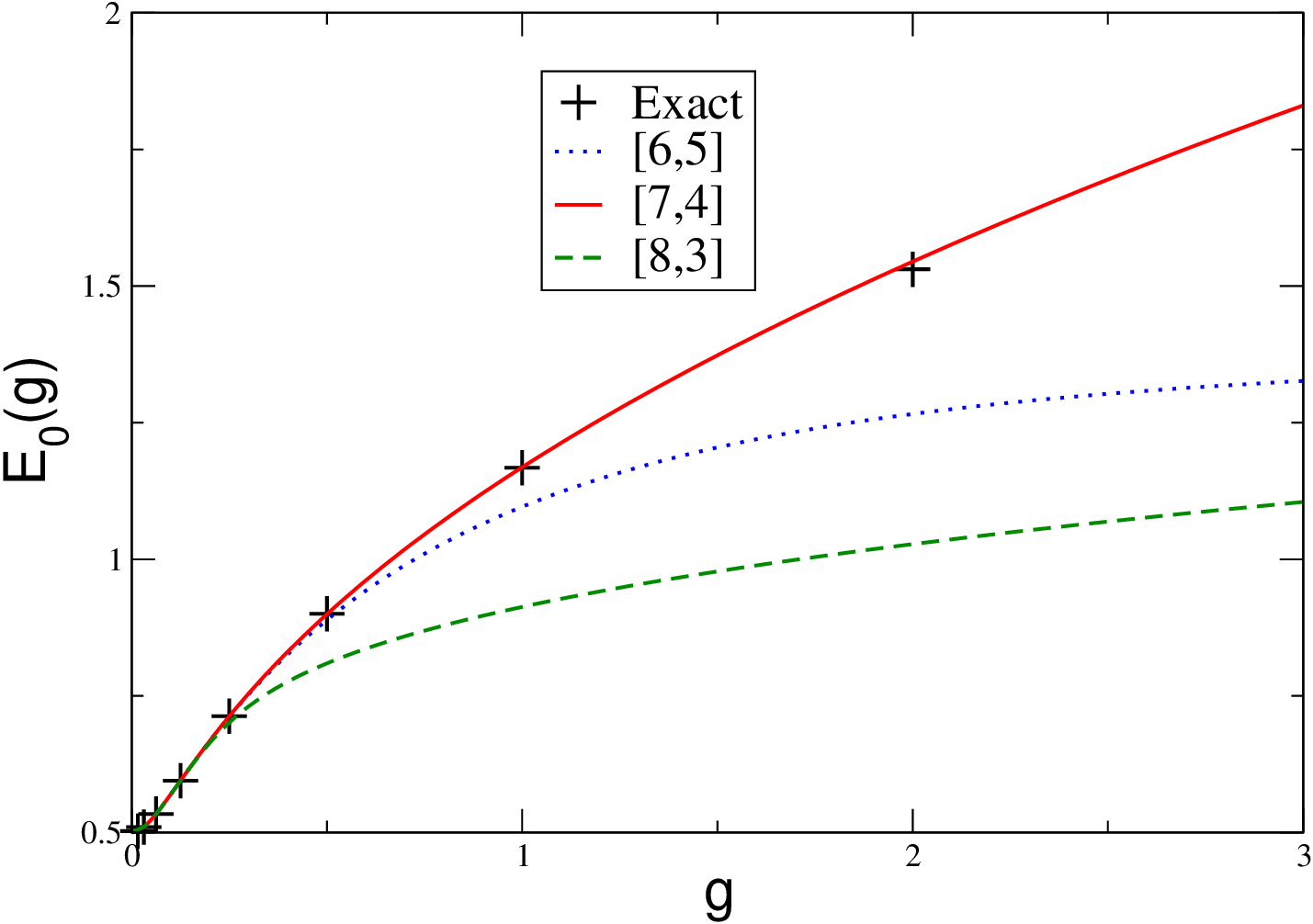}
\caption{(color online) Ground state energy of the anharmonic oscillator as a function
of $g$. The squares are numerical results, the curves correspond to the results obtained
with PPT to different orders.}
\label{Fig_2b}
\end{center}
\end{figure}

\subsection{Zero dimensional $\phi^4$ theory}

Integrals of the form
\beq
E(g) = \int_{0}^{+\infty} e^{-x^2-g x^4} dx
\label{ex3_1}
\eeq
can be used as a model of a $\phi^4$ in zero dimensions~\cite{JZJ,BDJ93,BDJ94}. 
As for the case of higher dimensional field theories, the perturbative series 
for this model is divergent.
The authors of \cite{BDJ93,BDJ94} have proved that the Linear Delta Expansion (LDE) 
is able to deal with this problem and that it provides results which rapidly converge to the 
exact value.

The integral in eq.(\ref{ex3_1}) admits an exact analytical solution which is given by
\beq
E(g) = \frac{e^{\frac{1}{8 g}}}{4 \sqrt{g}} \ K_{1/4}\left(\frac{1}{8 g}\right) \ ,
\label{ex3_2}
\eeq
where $K_{1/4}(g)$ is the Bessel function of order $1/4$. Notice that for negative values 
of $g$ this expression acquires an imaginary part, signaling that the system becomes metastable. 

We will now analyze this problem with the help of PPT. We choose the functional relation
\beq
g(\varrho) = \varrho \ \left[ \frac{1+\sum_{n=1}^{\bar{N}} c_n \varrho^{n}}{1+\sum_{n=1}^{\bar{N}+1} 
d_n \varrho^{n}}  \right]^5 ,
\label{ex3_3}
\eeq
which allows one to obtain the correct asymptotic behavior, $E(g) \propto g^{-1/4}$ as $g\rightarrow \infty$.

\begin{figure}
\begin{center}
\bigskip\bigskip\bigskip
\includegraphics[width=14cm]{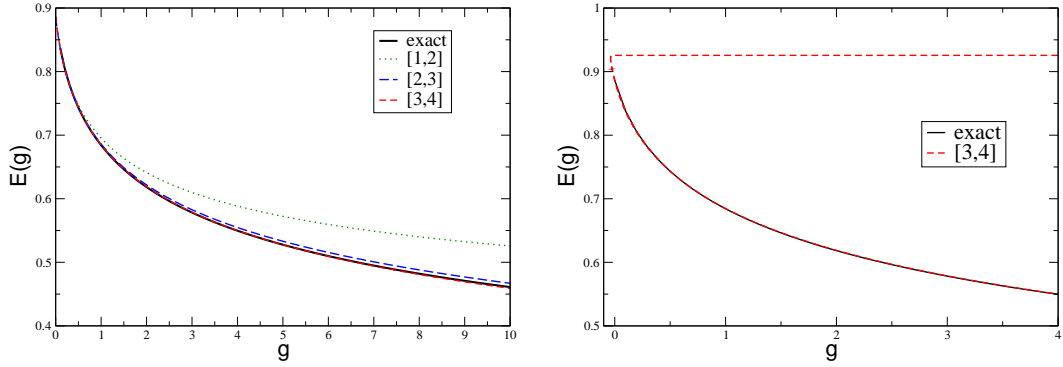}
\caption{(color online) 
Left panel: Comparison between the exact integral  for the zero dimensional $\phi^4$ theory and three different 
approximations obtained using the PPT; 
Right panel: comparison between the set $[3,4]$ and the exact integral. The approximate solution has a 
branch point close to $g=0$.}
\label{Fig_3}
\end{center}
\end{figure}

As usual the application of the PPT method requires that the coefficients $c_n$ and $d_n$ corresponding
to a given $\bar{N}$ be determined by imposing the PAS, i.e. by asking that the observable $E(g)$
be linear in the parameter $\varrho$. Using three different sets, corresponding to $\bar{N}=1,2$ and $3$
we have observed that our results converge quickly to the exact analytical result (see the left panel in
Fig.\ref{Fig_3}).

We will however move further and concentrate over the best set, $\bar{N} = 3$. In this case 
the relation between $g$ and $\varrho$ is given by
\beq
g(\varrho) \approx \frac{\rho  \left(2924.98 \varrho^3+ 881.78 \varrho^2+ 58.83 \varrho +1\right)^5}{\left(
-1737.20 \varrho^4+ 2243.93 \varrho^3+ 832.17 \varrho^2+ 57.96\varrho +1\right)^5} \ .
\label{ex3_4}
\eeq

Since $\varrho_0 = - 0.0593$ is a zero of the denominator, $\lim_{\varrho \rightarrow \varrho_0} g(\varrho) = \infty$:
this result signals the presence of a branch point in the proximity of $\varrho_0$ 
(see the right panel of Fig.\ref{Fig_3}).
We now consider the region $g < 0$, where the analytic continuation of the solution acquires an imaginary
part. Using eq.(\ref{ex3_4}) we find the numerical solutions of the equation $g(\varrho) = g$, with $g<0$.
For example, corresponding to $g=-1$ we find two pairs of complex conjugated roots accompanied by a single
real root:
\beq
\varrho_1 &=& 0.20784008231963882+0.5489736369789899 \ i \\
\varrho_2 &=& 0.20784008231963882-0.5489736369789899 \ i \\
\varrho_3 &=& -0.24838105054544726 \\
\varrho_4 &=& -0.2232185118896695+0.006357863317100693 i \\
\varrho_5 &=& -0.2232185118896695-0.006357863317100693 i \ .
\eeq

It is important at this point to notice that obtaining a complex value for $\varrho$ has an immediate 
effect on the observable $E(g)$, which {\sl acquires an imaginary part}, $Im E(g) = b_1 \ Im \varrho$.
We can verify if one of these solutions corresponds to the exact solution of (\ref{ex3_1}) for $g=-1$:
\beq
Im E(g) = -0.3767931291206198  
\eeq
which should be compared to the imaginary parts calculated with the PPT using the numerical roots $\varrho_i$, 
$i=1,\dots, 5$:
\beq
Im E(g)^{PPT}_1 &=& -0.36488641384088155 \\
Im E(g)^{PPT}_2 &=& 0.36488641384088155 \\
Im E(g)^{PPT}_3 &=& 0 \\
Im E(g)^{PPT}_4 &=& -0.004225882244972266 \\ 
Im E(g)^{PPT}_5 &=& 0.004225882244972266 .
\eeq

These results suggest that the first root corresponds to the analytic continuation of the solution for $g>0$
to negative values. On the other hand the real part of $Re E(g)^{PPT}_1 = 0.7480818175977717$ has the opposite
sign of $Re  E(g) = -0.7603309714715291$: this happens because our function is continous and therefore 
it is not possible to reproduce a discontinuity at $g=0$.

\begin{figure}
\begin{center}
\bigskip\bigskip\bigskip
\includegraphics[width=7cm]{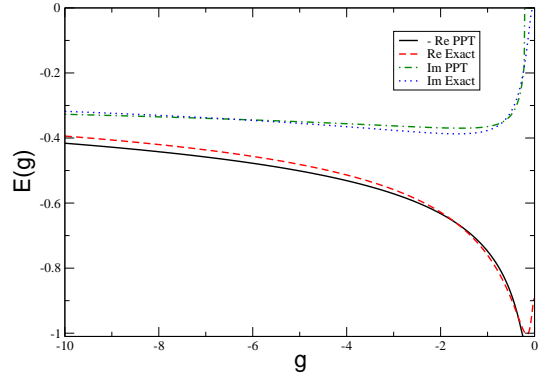}
\caption{(color online) 
Real and imaginary parts of $E(g)$ for the zero dimensional $\phi^4$ theory obtained using the PPT with $\bar{N} = 3$. 
The results are compared with the exact expression of eq.(\ref{ex3_1}).}
\label{Fig_3a}
\end{center}
\end{figure}

To test the conclusions that we have just reached we can plot the real and imaginary parts of $E(g)$ as obtained
from (\ref{ex3_1}) and compare them with the results provided by the PPT. In Fig.\ref{Fig_3a} we show the
results obtained with this comparison. We conclude that both the imaginary and real parts of $E(g)$ 
(apart for a sign) are reproduced with good quality; clearly the exponential  behavior of 
the exact solution for $g\rightarrow 0^-$ cannot  be reproduced in this approach.

\begin{figure}
\begin{center}
\bigskip\bigskip\bigskip
\includegraphics[width=9cm]{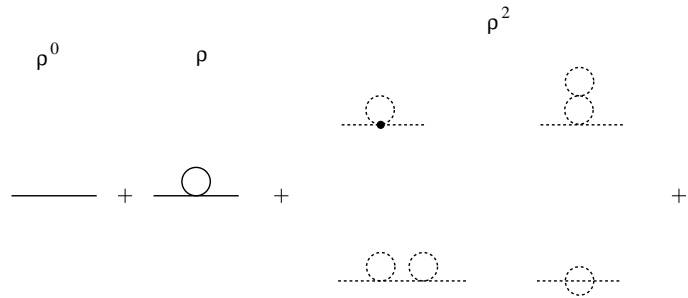}
\caption{Feynman diagrams for the propagator in a  $\phi^4$ theory. The diagrams with dashed line need to cancel fixing
the unknown coupling contained in the bold vertex. The coulings coming from the expansion of $g$ to order $\varrho^2$
and higher are represented with a bold circle.}
\label{Fig_3b}
\end{center}
\end{figure}

Let us now briefly explore a different issue. If we push forward the analogy with quantum field theory, 
the application of the PPT to this problem can have a simple interpretation in terms of Feynman diagrams: 
because applying the PAS we are expanding not in the coupling, but in a parameter $\varrho$, there is a 
infinite number of vertices appearing at tree level, whose couplings contain the unknown constants $c_n$ and $d_n$. 
The spirit of the PAS is then to perform a perturbative (in $\varrho$) calculation in which all diagrams 
corresponding to orders higher than $\varrho$ cancel out by fixing the unknown constants and therefore 
yielding the observable in terms of just two diagrams, with zero and one vertex respectively.  
This argument is sketched in Fig.\ref{Fig_3b}.

\subsection{QED effective action}

We consider the QED effective action in the presence of a constant background magnetic field:
\beq
S = - \frac{e^2 B^2}{8 \pi^2} \int_0^\infty \frac{ds}{s^2} \left\{ \coth s - \frac{1}{s} - \frac{s}{3} 
\right\} \ e^{-\frac{m_e^2 s}{e B}} \ ,
\eeq
where $B$ is the magnetic field strength, $e$ the electron charge and $m_e$ the electron mass. 
Following \cite{Jent} we introduce the effective coupling $g = e^2 B^2/m_e^2$ and obtain 
the divergent perturbative series
\beq
S =  - 2 \frac{e^2 B^2}{\pi^2} g \ \sum_{n=0}^\infty b_n g^n 
\eeq
with
\beq
b_n = (-1)^{n+1} \frac{4^n |{\cal B}_{2n+4}|}{(2 n+4) (2 n+3)  (2n+2)}  \ , 
\eeq
${\cal B}_{2n+4}$ being Bernoulli numbers.

\begin{figure}
\begin{center}
\bigskip\bigskip\bigskip
\includegraphics[width=7cm]{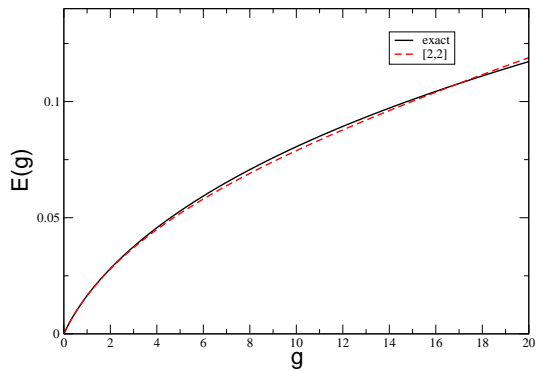}
\caption{(Color online) Comparison between the exact numerical result for $E(g) \equiv S(g)/(2 \frac{e^2 B^2}{\pi^2})$ and the
approximation obtained with the set $[4,4]$.}
\label{Fig_qed}
\end{center}
\end{figure}

In Fig.\ref{Fig_qed} we show the comparison between the exact numerical result 
for $E(g) \equiv S(g)/(2 \frac{e^2 B^2}{\pi^2})$ and the approximation obtained 
with the set $[4,4]$: although we have used only ten perturbative coefficients, 
the resummation provides a quite precise approximation over a large range of the
coupling in an analytical form.

\subsection{One plaquette integral}

In \cite{LiMeu} the weak coupling expansion for a one-plaquette SU(2) lattice gauge theory was discussed.
The partition function in this case is given by
\beq
Z(\beta) = \frac{2}{\pi} \int_{-1}^{+1} \sqrt{1-u^2} e^{-\beta (1-u)} \ du
\eeq
and can be calculated exactly in terms of the modified Bessel function $I_1$:
\beq
Z(\beta) = 2 e^{-\beta} \frac{I_1(\beta)}{\beta}
\eeq
This expression admits a convergent series expansion around $\beta=0$ (strong coupling expansion), 
but provides a divergent series when expanded in the opposite regime, $\beta \rightarrow \infty$, 
 (weak coupling expansion)\cite{LiMeu}:
 \beq
 Z(\beta) \approx (\beta \pi)^{-3/2}  2^{1/2} \sum_{l=0}^\infty \frac{(\Gamma(l+1/2))^2 (l+1/2)}{l! (1/2-l)}  \ . 
 \eeq
 The terms of this series grow like $l!/2^l$ and the sign does not oscillate.

 We will now apply our method to this model, considering both regimes and using as usual the functional relation
 \beq
 g(\varrho) = \varrho \ \left[ \frac{1+\sum_{n=1}^{\bar{N}_u} c_n \varrho^{n}}{1+\sum_{n=1}^{\bar{N}_d} 
 d_n \varrho^{n}}  \right] .
 \eeq
to be determined independently in the two regimes.

 \subsubsection{Weak coupling expansion}

 In this case we identify $\beta = 1/g$ and use  a set corresponding to $\bar{N}_u = 2$ and $\bar{N}_d = 3$, 
obtaining the solution
\beq
\frac{1}{\beta} = g(\varrho) = \frac{\varrho  \left(1.816 \varrho^2- 3.428 \varrho + 1\right)}{ 0.154 \varrho^3+
0.920 \varrho^2- 3.115 \rho +1}
\eeq

 \subsubsection{Strong coupling expansion}

 In this case we identify $\beta = g$ and 
a set corresponding to $\bar{N}_u = 6$ and $\bar{N}_d = 3$, obtaining the solution
 \beq
 \beta = g(\varrho) \approx \frac{\varrho  \left(-0.0000545\varrho^6-0.000489 \varrho^5-
0.00349 \varrho^4- 0.0324 \varrho^3+ 0.636 \varrho^2- 1.555 \varrho +1\right)}{-0.327 \varrho^3+
1.509 \varrho^2- 2.180 \varrho +1}
\eeq

\begin{figure}
\begin{center}
\bigskip\bigskip\bigskip
\includegraphics[width=7cm]{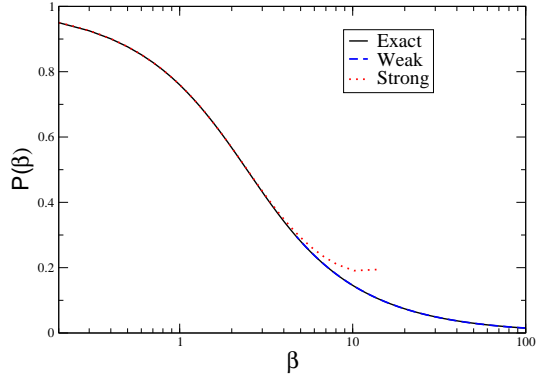}
\caption{(color online) $P$ versus $\beta$ for SU(2) on one plaquette. The solid line is the exact
result; the dashed line corresponds to the Weak Coupling Expansion; the dotted line corresponds to the
Strong Coupling Expansion. }
\label{Fig_su2}
\end{center}
\end{figure}

In Fig.\ref{Fig_su2} we have plotted $P =  - \frac{d}{d\beta} ln Z$ as a function of $\beta$,
as done also in Fig.3 of \cite{LiMeu}. Our results show that both the Weak and Strong coupling 
expansions, resummed through the PPT converge to the exact result: this is particularly remarkable
in the case of the weak coupling expansion.

\subsection{$\phi^4$ field theory in $1+1$}

In a recent paper Nishiyama~\cite{Nish01} has studied a lattice $\phi^4$ model in $1+1$ dimensions, described by
the hamiltonian
\beq
\hat{H} =  \sum_i \left[ \frac{1}{2} \hat{\pi}_i^2 + \hat{\phi}_i^4 + g \left( \frac{1}{2} 
\left(\hat{\phi}_i-\hat{\phi}_{i+1}\right)^2+ \frac{1}{2} \hat{\phi}_i^2 \right)\right] \ ,
\eeq
where $i$ is the site index and $\hat{\pi}_i$ and $\hat{\phi}_i$ are canonically conjugated operators. Notice that
we have changed the notation in \cite{Nish01} adopting the conventions used in this paper.

Using a linked cluster expansion Nishiyama has obtained the perturbation series in $g$ up to order 11:
\begin{eqnarray}
E(g) &=& 0.66798625915577710827096201688  +0.43100635014259473006095738275 g \nonumber \\
     &-&0.10148809521111863294125944502 g^2 +0.04803845646443637442034775341 g^3 \nonumber \\
     &-&0.029018513979643624653232757064 g^4 +0.019777791330895673863274529570  g^5 \nonumber \\
     &-&0.014454753622894705466341917665 g^6  +0.01106139124598227911409431586 g^7 \nonumber \\
     &-&0.0087493465269972 g^8 +0.007096747591805 g^9 
     - 0.005871428 g^{10}+0.00493622 g^{11}  \ .
\label{series_expansion}
\end{eqnarray}

Since the perturbative series has a radius of convergence $g_0 \approx 1$, an Aitken $\delta^2$ 
process was used in \cite{Nish01} to accelerate the convergence of this series. The accelerated
series was then compared with the numerical results obtained using the Density Matrix Renormalization
Group (DMRG), showing that the region of convergence could be enlarged up to $g \approx 2$.

\begin{figure}
\begin{center}
\bigskip\bigskip\bigskip
\includegraphics[width=7cm]{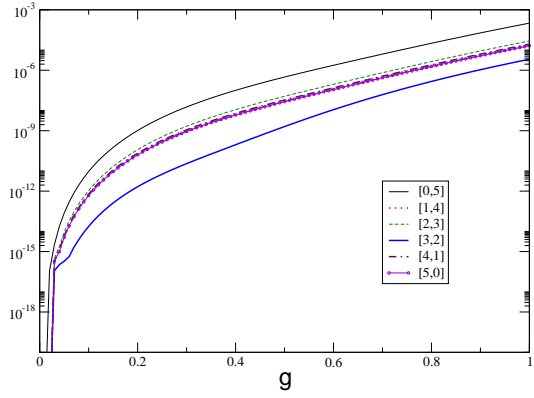}
\caption{(color online) 
Difference between the perturbative polynomial of order $g^{11}$ given by Nishiyama and the polynomial
of order $g^{11}$ constructed with PPT working to order $N_u + N_d = 5$.}
\label{Fig_4}
\end{center}
\end{figure}

We will now apply our method to this problem and consider
\beq
g(\varrho) = \varrho \ \left[ \frac{1+\sum_{n=1}^{\bar{N}_u} c_n \varrho^{n}}{1+\sum_{n=1}^{\bar{N}_d} 
d_n \varrho^{n}}  \right] ,
\eeq
where as usual $N_u$ and $N_d$ fix the asymptotic behavior for $g\rightarrow \infty$. As we do not know
this behavior exactly we will work at order $N_u + N_d = 5$ and take into account all the possible combinations
of $\bar{N}_u$ and $\bar{N}_d$ keeping the sum fixed.
In this case only the coefficients $b_n$ with $n$ going from 0 to 6 are used, the remaining being a prediction
of our method. In Fig.\ref{Fig_4} we have plotted the difference between the perturbative polynomial of 
order $g^{11}$ given by Nishiyama and the polynomial of order $g^{11}$ constructed with PPT working to order
$N_u + N_d = 5$. As one can see, the set $[3,2]$ provides the smallest difference, a result which suggests 
the asymptotic behavior of the energy as $E \propto \sqrt{g}$ for $g\rightarrow \infty$.
 
In Table \ref{table3} we have compared the exact coefficients calculated by Nishiyama with those predicted 
by the PPT working with the set $[3,2]$. The last row of this table shows the errors  
$\Sigma_n \equiv 100 \times \left| b_n^{[3,2]}/b_n^{exact}-1 \right|$: from this results we can conclude
that resummation through PPT allows to achieve a truly remarkable precision, the largest error being of 
about $0.1  \%$.

\begin{widetext}
\begin{table}
\caption[t1]{Comparison between the perturbative coefficients of \cite{Nish01} and those predicted with
PPT working with the set $[3,2]$.}
\begin{tabular}{|c|c|c|c|c|c|}
\colrule
              & $b_7$  & $b_8$ & $b_9$ & $b_{10}$ &  $b_{11}$ \\
\colrule
$b_n^{exact}$ & 0.011061391245982 & -0.0087493465269972 & 0.007096747591805 & -0.005871428 & 0.00493622 \\
$b_n^{[3,2]}$ & 0.011061133480144 & -0.0087483769472128 & 0.007094602847397 &  -0.00586767 & 0.00493037 \\
error ($\%$) & 0.00233 & 0.011081739435 & 0.03022 & 0.06403 & 0.11851 \\
\colrule
\end{tabular}
\label{table3}
\end{table}
\end{widetext}

\begin{figure}
\begin{center}
\bigskip\bigskip\bigskip
\includegraphics[width=7cm]{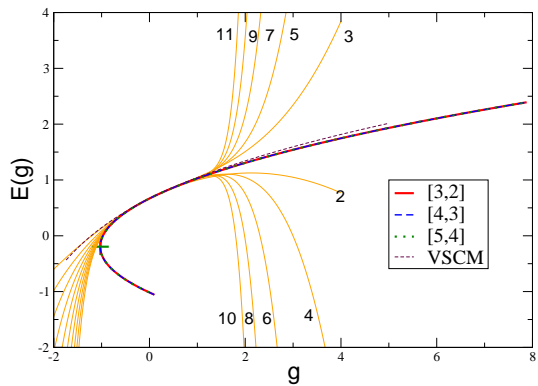}
\caption{(color online) 
Comparison between the resummed energies to orders $[3,2]$, $[4,3]$ and $[5,4]$ and the perturbative 
polynomials.}
\label{Fig_5}
\end{center}
\end{figure}

In Fig.\ref{Fig_5} we have compared the perturbative polynomials for the energy from orders $g^2$ to $g^{11}$ with the 
energy resummed with the sets $[3,2]$, $[4,3]$ and $[5,4]$. There are several striking aspects which should 
impress the reader: first of all, the difference bewteen the three sets is extremely thiny, thus signaling that the
convergence is extremely strong; in second place, the resummed energy confirms the DMRG result displayed in
Fig.2 of \cite{Nish01}; finally, the resummed energy is a multivalued function, with a branch point at 
$g \approx -1.025$. This last finding is extremely interesting, because in \cite{Nish01} it was speculated
the existence of a phase transition at $g\approx -2$ (in our notation): the resummed energy plotted in Fig.2
of \cite{Nish01} appears to have a singularity around $g = -1$ (in our notation), i.e. in the same region where 
we observe the branch point~\footnote{Clearly, the branch point of function $y = f(x)$ at a point $x=x_0$ 
manifests itself as a singularity in that point when it is calculated using the Taylor series around a 
different point.}.
Because of the use of a parameter $\varrho$, PPT can deal with multivalued functions in a way which is not 
possible in conventional perturbation theory. 
Finally, the thiny dashed line in the plot corresponds to the numerical result obtained in \cite{Amore06a} using
the Variational Sinc Collocation Method (VSCM) within a mean field approach.

\subsection{Elastic theory}

Another example of divergent series has been studied in \cite{Buch96}. The authors of that paper have found out 
that the series for the inverse bulk modulus $K$ as a function of the compression:
\beq
\frac{1}{K} = b_0 + b_1 P + b_2 P^2 + \dots
\eeq
has zero radius of convergence and they obtained an explicit expression for the coefficients in a simplified 
calculation. In the following we will uniform the notation to the conventions used in this paper and 
refer to the pressure $P$ as the coupling $g$.
The coefficients of the series are~\cite{Buch96}
\beq
b_n = -(n+2) \frac{f_{n+2}}{A}
\eeq
where
\beq
f_n = (-1)^{n+1} \ \Gamma\left(\frac{n+1}{2}\right)  \ \left( \frac{\pi \sqrt{1-\sigma^2}}{4 \beta Y 
\alpha^2}\right)^{n/2} \ \left(\frac{2\pi A}{\lambda^2}\right) \ \frac{\sqrt{1 - \sigma^2}}{2 \sqrt{\pi} 
\beta^{5/2} \alpha \lambda^2 \sqrt{Y}}  \ .
\eeq

Here $\alpha$ is the surface tension, $Y$ is the Young's modulus, $\sigma$ is the Poisson ratio, 
$\beta = 1/k_B T$ and $\lambda$  is the ultraviolet cutoff of the theory. To apply our method we introduce
\beq
g(\varrho) = \varrho \  \frac{1+\sum_{n=1}^{\bar{N}_u} c_n \varrho^{n}}{1+\sum_{n=1}^{\bar{N}_d}  d_n \varrho^{n}}  
\eeq
and fix $\bar{N}_u+\bar{N}_d = 5$. Working to this order the first predicted coefficient is $b_7$.
We have performed a calculation using $\beta = \lambda = \alpha = Y = 1$ and $\sigma = 1/2$.

\begin{widetext}
\begin{table}
\caption[t1]{Comparison of the exact perturbative coefficient $b_7$ for the elastic series with the approximate 
coefficient predicted by PPT using $\bar{N}_u +\bar{N}_d = 5$. $b_7^{exact} = -\frac{729 \pi ^5}{8192} 
\approx -27.2324646250573949498$.
The parameters are chosen $\beta = \lambda = \alpha = Y = 1$ and $\sigma = 1/2$. }
\begin{tabular}{|c|c|c|c|c|c|c|}
\colrule
      & $b_7^{[5,0]}$  & $b_7^{[4,1]}$ & $b_7^{[3,2]}$ & $b_7^{[2,3]}$ &  $b_7^{[1,4]}$ & $b_7^{[0,5]}$ \\
\colrule
$b_7$ & -48.40832399 & -27.41931538  & -27.18605118 & -27.21965479 & -27.17382665 & -27.46859099 \\
error ($\%$) & $77.75$ & $0.69$ & $0.17$ & $0.047$ & $0.21$ & $0.87$ \\
\colrule
\end{tabular}
\label{table4}
\end{table}
\end{widetext}

As we have seen from Table \ref{table4} the optimal set for $\bar{N}_u+\bar{N}_d$ is $[2,3]$, corresponding to an asymptotic behavior
$1/K \propto g^0$. At this order we have found~\footnote{Although the coefficients $c_n$ and $d_n$ are calculated
exactly, we prefer to write them numerically to allow a more compact expression.}:
\beq
g(\varrho) = \frac{\varrho  \left(-0.4923950887 \varrho^2- 0.94207053700 \varrho + 1\right)}{
0.6331671415\varrho^3+ 0.9634343699 \varrho^2-2.4724637506 \varrho +1}
\eeq

Working with to order $\bar{N}_u+\bar{N}_d=7$ we have found that the best set is the $[5,2]$, which provides
a different asymptotic behavior for the inverse compression modulus, $1/K \propto g^{1/4}$.
In Table \ref{table5} we compare the predictions for the perturbative coefficients obtained with the two different 
sets: notice that the second set does not predict the coefficients $b_7$ and $b_8$. The second set
gives more precise results than the first set for $b_9$ and $b_{10}$, but a slightly worse error
for $b_{11}$.

\begin{figure}
\begin{center}
\bigskip\bigskip\bigskip
\includegraphics[width=7cm]{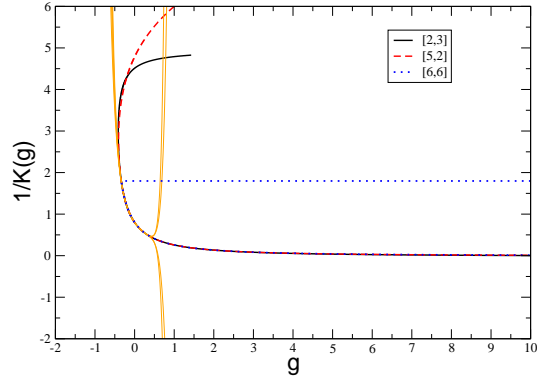}
\caption{(color online) Comparison between the perturbative polynomials of order $8$ through $10$ and the results 
obtained using the sets $[2,3]$ and $[5,2]$ with PPT .}
\label{Fig_6}
\end{center}
\end{figure}

In Fig.\ref{Fig_6} we have compared the perturbative polynomials of order $8$ through $10$ with PPT results
corresponding to the sets $[2,3]$,$[5,2]$ and $[6,6]$: just as in the case of the lattice $\phi^4$ previously discussed
we observe a branch point in the resummed solution, corresponding to $g_0^{[2,3]} \approx -0.412879$, 
$g_0^{[5,2]} \approx -0.401549$ and $g_0^{[6,6]} \approx -0.330171$ with the different sets. 

If we go back to the example of $\phi^4$ in zero dimensions, there we have seen that $g=0$ is a point where 
the function is not analytical and therefore the perturbative series is divergent. In the present example we 
can use the words of the authors of \cite{Buch96} and say that ``under stretching ($g<0$) the true ground state 
is fractured into pieces. As a result $g=0$ cannot be a point of analyticity for $K(g)$ and thus the series 
has zero radius of convergence''. 
Our results however display a branch point not exactly at $g=0$ or close to it as in the case of the 
$\phi^4$ model in zero dimensions.

\begin{widetext}
\begin{table}
\caption[t1]{Comparison between the perturbative coefficients of \cite{Buch96} and those predicted with
PPT working with the set $[2,3]$. We use  $\beta = \lambda = \alpha = Y = 1$ and $\sigma = 1/2$.
Coefficients with a dagger are input of the method.}
\begin{tabular}{|c|c|c|c|c|c|}
\colrule
              & $b_7$  & $b_8$ & $b_9$ & $b_{10}$ &  $b_{11}$ \\
\colrule
$b_n^{exact}$ & - 27.2324646250574 & 51.2814820354647 & - 100.257786099872 & 203.060091610056 & -425.208317773323 \\
\colrule
$b_n^{[2,3]}$ & - 27.2196547938467 & 51.1255824090448 & - 99.224725761445  & 197.961957448658 & -561.00565159656  \\
error ($\%$) & 0.047 & 0.304 & 1.03 & 2.51 & 31.9 \\
\colrule
$b_n^{[5,2]}$ &  - 27.2324646250574$^\dagger$ & 51.2814820354647$^\dagger$ & -100.216793832003 & 202.536703574420 & -579.771361475677 \\
error ($\%$) &  0 &  0 & 0.041 & 0.26 & 36.35 \\
\colrule
\end{tabular}
\label{table5}
\end{table}
\end{widetext}

\subsection{Elliptic integral of the first kind}

Consider the elliptic integral of the first kind:
\beq
E(g) = K(g) \equiv \int_0^{\pi/2} \frac{dt}{\sqrt{1- g \sin^2 t}} \ ,
\eeq
which diverges for $g \rightarrow 1$. It also obeys the series representation 
\beq
K(g) = \frac{\pi}{2}  \sum_{k=0}^\infty \frac{\left(\frac{1}{2}\right)_k\left(\frac{1}{2}\right)_k  }{k!^2} g^k 
\label{ellip}
\eeq
which converges for $|g|<1$. 

We want to show that it is possible to resum the perturbative series using the PPT. We choose
the functional form:
\beq
g(\varrho) = \varrho \  \frac{1+\sum_{n=1}^{\bar{N}} c_n \varrho^{n}}{1+\sum_{n=1}^{\bar{N}+1}  d_n \varrho^{n}}  \ .
\eeq
The choice of $\bar{N}_u +1 = \bar{N}_d$ is not arbitrary: with this choice we have that $\lim_{\varrho\rightarrow \infty} g(\varrho) = \bar{g} < \infty$,
which means that the resummed function will have a singularity precisely at $g=\bar{g}$.

Using $\bar{N}=2$ we find 
\beq
g(\varrho) = \frac{\varrho  \left(\frac{381 \varrho ^2}{35840}+\frac{187 \varrho}{2240}+1\right)}{\frac{2301 \varrho ^3}{286720}+
\frac{1181 \varrho^2}{8960}+\frac{1447 \varrho }{2240}+1}
\eeq
which predicts the singularity of the elliptic integral at
\beq
\bar{g}^{[2,3]} = \frac{1016}{767} \approx 1.324  \ .
\eeq

Increasing $\bar{N}$ this singularity moves towards its exact value, $\bar{g}=1$; for example, using $\bar{N} = 3,4$ and $5$ and find 
\beq
\bar{g}^{[3,4]} = \frac{4999}{3752} \approx 1.332 \ , \ 
\bar{g}^{[4,5]} =\frac{5509}{8216} \approx 0.670 \ , \
\bar{g}^{[5,6]} = \frac{69944792}{67596985} \approx 1.035 \ .
\eeq

\begin{figure}
\begin{center}
\bigskip\bigskip\bigskip
\includegraphics[width=7cm]{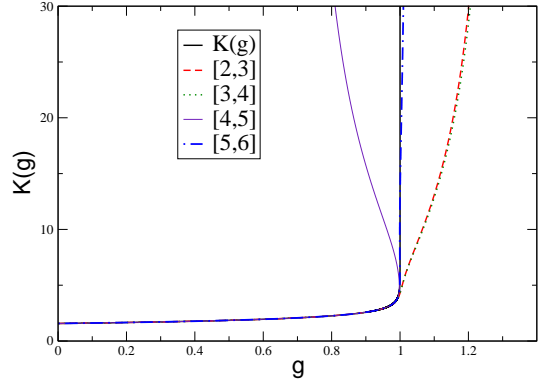}
\caption{(color online) Comparison between elliptic integral $K(g)$ and the PPT approximations obtained with sets 
$[2,3]$, $[3,4]$, $[4,5]$ and $[5,6]$.}
\label{Fig_7}
\end{center}
\end{figure}

The reader will notice that the singularity predicted by the set $[4,5]$ falls below the exact singularity: the reason for this 
behavior is easily understood looking at Fig.\ref{Fig_7}. As a matter of fact the set $[4,5]$ (the thin line in the plot)
has a branch point close to $g=1$, and therefore the singularity belongs to the nonphysical branch. Notice that the set 
$[5,6]$ provides an excellent approximation.

Let us now compare the expansion of $K(g)^{[2,3]}$ around $g=0$ with the exact result, provided by the series (\ref{ellip}).
We have
\beq
K(g)^{[2,3]} &\approx& 
\frac{\pi }{2}+\frac{\pi g}{8}+\frac{9 \pi  g^2}{128}+\frac{25 \pi  g^3}{512}+\frac{1225 \pi  g^4}{32768}+\frac{3969 \pi  g^5}{131072}+
\frac{53361 \pi  g^6}{2097152}+\frac{206126367 \pi  g^7}{9395240960}\nonumber \\
&+&\frac{405813405891 \pi  g^8}{21045339750400}+
\frac{810831328918663 \pi g^9}{47141561040896000}+\frac{1639189758117069059 \pi g^{10}}{105597096731607040000} \nonumber \\
&+& \frac{3345592829494380888687 \pi  g^{11}}{236537496678799769600000}+\frac{6882636481373124653844491 \pi  g^{12}}{529843992560511483904000000}+\dots \nonumber \\
&\approx& 1.5707963267949 + 0.392699081698724 g + 0.220893233455532 g^2 +  0.153398078788564 g^3 \nonumber \\
&+& 0.117445404072494 g^4 + 0.09513077729872 g^5 + 0.079936278146842 g^6 + 0.06892479746239 g^7 \nonumber \\
&+& 0.060578751866013 g^8 + 0.054035158997418 g^9 + 0.0487671220263656 g^{10} \nonumber \\
&+& 0.0444347725101475 g^{11} + 0.0408090692936214 g^{12}  + \dots
\eeq
and
\beq
K(g) &\approx& 
\frac{\pi }{2}+\frac{\pi g}{8}+\frac{9 \pi  g^2}{128}+\frac{25 \pi  g^3}{512}+\frac{1225 \pi  g^4}{32768}+\frac{3969 \pi  g^5}{131072}+
\frac{53361 \pi  g^6}{2097152}+\frac{184041 \pi  g^7}{8388608} \nonumber \\
&+& \frac{41409225 \pi  g^8}{2147483648}+\frac{147744025 \pi  g^9}{8589934592}+\frac{2133423721 \pi  g^{10}}{137438953472}+
\frac{7775536041 \pi  g^{11}}{549755813888}+\frac{457028729521 \pi  g^{12}}{35184372088832}+\dots \nonumber \\
&\approx& 1.5707963267949 + 0.392699081698724 g + 0.220893233455532 g^2 + 0.153398078788564 g^3  \nonumber \\
&+& 0.117445404072494 g^4 + 0.0951307772987205 g^5 + 0.0799362781468415 g^6 + 0.0689246479939603 g^7 \nonumber \\
&+& 0.0605783039009417 g^8 + 0.0540343513190498 g^9 + 0.0487660020654424 g^{10} \nonumber \\
&+& 0.0444334853530168 g^{11} + 0.0408078363745588 g^{12} + \dots \ .
\eeq

Notice that coefficients starting from $7$ and higher, are {\sl predictions} of the PPT: we see, for example, that the 
coefficient of the term of order $g^{12}$ is predicted with an error $0.003 \%$!

\subsection{Virial coefficients}

As a last example we consider the low density virial expansion of the pressure
\beq
\frac{P}{k_B T} = \sum_{n=1}^\infty B_k \rho^k \ ,
\label{vc_1}
\eeq
where the $B_k$ are the virial coefficients and $\rho$ is the density (not to be confused
with $\varrho$). Table 1 of \cite{Clis06} contains the numerical values of the virial coefficients
for hard spheres in $D$ dimensions, with $2\leq D\leq 8$. In the following we will uniform 
the notation  in (\ref{vc_1}) to the notation adopted in this paper and call $g$ the density.

As usual we adopt the functional form
\beq
g(\varrho) = \varrho \  \frac{1+\sum_{n=1}^{\bar{N}_u} c_n \varrho^{n}}{1+\sum_{n=1}^{\bar{N}_d}  d_n \varrho^{n}}  \ .
\eeq

In Table \ref{table6} we have used PPT with $\bar{N}_u+\bar{N_d}=5$ to predict the virial coefficients
$B_9$ and $B_{10}$ from the previous one. As one can see the set $[2,3]$ provides highly precise results.
We have then used the set $[3,4]$ which has the same behaviour for $\varrho \rightarrow\infty$ to obtain
an estimate for the eleventh virial coefficient. To orders $[2,3]$ and $[3,4]$ we have found
\beq
\left[B_{11}/B_2^{10}\right]^{[2,3]} = 0.01094432  \ \ \ , \ \ \ \left[B_{11}/B_2^{10}\right]^{[3,4]} = 0.01090061 \ .
\eeq

Since these results are not (yet) available in the literature, the prediction made here will be a strong 
test of the present method once the calculation of $B_{11}$ will be made.
In Table \ref{table7} we  compare our predictions for the virial coefficients going
from $B_{11}$ to $B_{18}$ with the predictions made in \cite{Clis06}. Our predictions are very close to those 
made by Clisby and McCoy for $D=2$.

Notice that finding  $N_u +1 = N_d$ has an important effect: as discussed in the previous example of the elliptic integral, 
in this case the solution will have a singularity at a finite value of $g$. For example, if we consider the set
$[3,4]$ the tranformation reads
\beq
g(\varrho) = \frac{\varrho  \left( 0.3004 \varrho^3+ 1.8876 \varrho^2+ 2.7141 \varrho + 1\right)}{0.2584 \varrho^4+
2.1266 \varrho^3+ 4.6986 \varrho^2+ 3.7831 \varrho +1}
\eeq
and the singularity is predicted to be fall at
\beq
\bar{g}^{[3,4]} &=& \lim_{\varrho \rightarrow \infty} g(\varrho) \approx 1.1625 .
\eeq

We would like to stress that this singularity is ``physical'', i.e. it is a singularity of the resummed function, 
in contrast with the singularity falling at the radius of convergence of a series. 
We can use the previous example of the elliptic integral to better understand this point:
in that case the perturbative series around $g=0$ had a radius of convergence $1$, coinciding with 
the location of the true singularity of $K(g)$. 

If we trust our result, we may conclude that the $\frac{P}{k_B T}$ becomes infinite at a finite density 
$g \approx \bar{g}^{[3,4]}$.

We have also considered the virial series in $D=3$ dimensions. Also in this case we have found out that, working with $\bar{N}_u+\bar{N}_d=5$
the optimal set corresponds to $[2,3]$ (see Table \ref{table6}) and therefore the virial series is expected to have a singularity at finite
density:
\beq
\bar{g}^{[3,4]} &=& \lim_{\varrho \rightarrow \infty} g(\varrho) \approx 1.43439 \ .
\eeq

In Table \ref{table7} we have also compared our predictions obtained with the set $[3,4]$ for $D=3$ with those made in \cite{Clis06}.
Unlike in the previous case, our result agree to some extent with those of \cite{Clis06} only for the coefficient $B_{11}$, whereas 
completely different predictions are made for the remaining coefficients.

\begin{table}
\caption[t1]{Virial coefficients for a hard spheres in $2$ and $3$ 
dimensions given in Table I of \cite{Clis06} and predictions 
using PPT with different sets.}
\begin{tabular}{|c|c|c|c|c|c|}
\colrule
\multicolumn{1}{|c|}{} 
&\multicolumn{1}{c}{$B_9/B_2^8$}
&\multicolumn{1}{c|}{$B_{10}/B_2^9$}
&\multicolumn{1}{c}{$B_9/B_2^8$}
&\multicolumn{1}{c|}{$B_{10}/B_2^9$} \\
\multicolumn{1}{|c|}{} 
&\multicolumn{2}{|c|}{$D=2$}
&\multicolumn{2}{c|}{$D=3$} \\
\colrule
Ref.\cite{Clis06}&         $0.0362193$  & $0.0199537$ & $0.0013094$ & $0.0004035$\\
\colrule
$[0,5]$ & $0.03739998$ & $0.02496595$ &  $0.0023400$ & $0.0031580$ \\
$[1,4]$ & $0.03625994$ & $0.02008503$ &  $0.0013509$ & $0.0004884$ \\
$[2,3]$ & $0.0362321$  & $0.0199843$  &  $0.0013165$ & $0.0004198$ \\
$[3,2]$ & $0.0362551$  & $0.02006717$ &  $0.0013404$ & $0.0004664$ \\
$[4,1]$ & $0.0368599$  & $0.02258546$ &  $0.0017325$ & $0.0014442$ \\ 
$[5,0]$ & $0.1747048$  & $0.75984885$ &  $0.0222648$ & $0.0735264$ \\
\colrule
\end{tabular}
\label{table6}
\end{table}

\begin{table}[H]
\centering
\caption{Predicted coefficients for approximants with 10 exact 
         coefficients for $D=2$ and $D=3$. Comparison between 
         the predictions of \cite{Clis06} and the 
         predictions obtained using the set $[3,4]$.}
\label{table7}
\vspace{2ex}
\scriptsize
\begin{tabular}{c|cccccccc}
\colrule
\\[-1.5ex]
  &\multicolumn{1}{c}{$B_{11}/B_2^{10}$} &\multicolumn{1}{c}{$B_{12}/B_2^{11}$} &\multicolumn{1}{c}{$B_{13}/B_2^{12}$} &\multicolumn{1}{c}{$B_{14}/B_2^{13}$} &\multicolumn{1}{c}{$B_{15}/B_2^{14}$} &\multicolumn{1}{c}{$B_{16}/B_2^{15}$} &\multicolumn{1}{c}{$B_{17}/B_2^{16}$} &\multicolumn{1}{c}{$B_{18}/B_2^{17}$}\\[0.7ex]
\colrule
\\[-1.5ex]
$D=2$  & & & & & & & \\
Ref.\cite{Clis06}  & $1.089 \times 10^{-2}$  & $5.90 \times 10^{-3}$ & $3.18 \times 10^{-3}$ & $1.70 \times 10^{-3}$ & $9.10 \times 10^{-4}$ 
                   & $4.84 \times 10^{-4}$   & $2.56 \times 10^{-4}$ & $1.36 \times 10^{-4}$ \\
$[3,4]$            & $1.0901 \times 10^{-2}$ & $5.9235 \times 10^{-3}$ & $3.2117 \times 10^{-3}$ & $1.7421 \times 10^{-3}$ &  $9.4698 \times 10^{-4}$ 
& $5.1638 \times 10^{-4}$ & $2.8247 \times 10^{-4}$  & $1.5492 \times 10^{-4}$ \\
\colrule
$D=3$  & & & & & & & \\
Ref.\cite{Clis06}  &$ 1.22 \times 10^{-4}$ &$ 3.64 \times 10^{-5}$ &$ 1.08 \times 10^{-5}$ &$ 3.2 \times 10^{-6}$ &$ 9.2 \times 10^{-7}$ &$ 2.6 \times 10^{-7}$ &&\\
$[3,4]$ & $1.1599 \times 10^{-4}$ & $2.2229 \times 10^{-5}$ & $-8.5616 \times 10^{-6}$ &  $-1.8088 \times 10^{-5}$ & 
$-2.0325 \times 10^{-5}$ & $-2.0112 \times 10^{-5}$ &  $-1.9136 \times 10^{-5}$ & $-1.7971 \times 10^{-5}$ \\
\colrule
\end{tabular}
\normalsize
\end{table}

\section{Conclusions}
\label{concl}

We have developed a new method, Parametric Perturbation Theory (PPT), which is alternative to the ordinary 
perturbation theory, i.e. does not amount to an expansion in any physical parameter. 
We have shown that PPT can used either as a fully autonomous perturbation scheme, as done in Section \ref{sec1}, 
or it can be applied to the coefficients of the perturbative expansion, resumming the series and providing 
physically meaningful results, as done in Section \ref{sec2}.
There are several aspects of our method which should make it very appealing:
\begin{itemize}
\item since PPT can use perturbative results as an input, it can be applied with limited effort to the huge amount of
problems which have been studied perturbatively; 
\item it provides analytical approximations;
\item unlike variational methods, such as the LDE or VPT, our method does not require any optimization in a 
variational parameter;
\item  although the asymptotic (strong coupling) behavior of the solution can be used, when known, to refine 
the functional relation $g=g(\varrho)$, PPT is capable of selecting the most appropriate asymptotic behavior 
of the solution within a class of different behaviors allowed to a given order; 
\item it predicts the unknown perturbative coefficients with high precision; 
\item it can easily describe multivalued functions and therefore is capable to produce
branch points at finite order, as observed in the examples: if these points are related 
to phase transitions of a system, as claimed in \cite{Nish01} in the case of $\phi^4$ in $1+1$,
then our method could provide an alternative tool to the study of critical phenomena; 
\item it can produce singularities in an observable working at finite order, as seen for the cases of the 
elliptic integral of first kind and for the virial coefficients of a hard sphere gas; 
\item most importantly, it can produce the nonperturbative imaginary part of an observable, which appears when a system
becomes metastable.
\end{itemize}

Future directions of work will certainly include the development of an autonomous perturbation scheme for quantum
mechanical problems, in analogy to the one developed in classical mechanics and the application of our results 
to resum perturbative calculations in quantum field theory. It will be also interesting to apply PPT to obtain 
new analytical approximations for special functions of relevance in Physics, as done in this paper with the elliptic
integral of the first kind.

\end{document}